\documentclass[%
 reprint,
 amsmath,amssymb,
 aps,
]{revtex4-2}

\usepackage{graphicx}
\usepackage{dcolumn}
\usepackage{bm}

\usepackage{mathrsfs}
\usepackage[scr=rsfs,cal=boondox]{mathalfa}

\usepackage{hyperref, color, xcolor}
\usepackage[normalem]{ulem}

\begin{document}

\title{Spontaneous Hole Formation in Cell Monolayers Emerges from Collective Cell Motion}

\author{Diogo E. P. Pinto}
\author{Jan Rozman}
\author{Julia M. Yeomans}%
 \email{julia.yeomans@physics.ox.ac.uk}
\affiliation{The Rudolf Peierls Centre for Theoretical Physics, Clarendon Laboratory, Parks Road, Oxford, OX1 3PU, UK}

\begin{abstract}

Although cell monolayers typically remain confluent, they can spontaneously develop persistent holes as a result of collective cellular motion. Recent studies on MDCK monolayers cultured on soft substrates have revealed that cells can align to create regions of local nematic order, and topological defects that generate localised mechanical stresses which can spontaneously trigger hole formation. To investigate this process, we develop a continuum multi-phase field model that incorporates internal dissipation and active dipolar forces that drive cell shape anisotropy. Our simulations show that reducing substrate friction enhances cell-cell velocity correlations. In this low-friction regime, topological defects give rise to spiral flow patterns that concentrate stress and can trigger hole formation. We further demonstrate that the number and stability of the holes—whether they close or persist—depends on both substrate friction and cellular activity. These findings underscore the critical role of collective cell dynamics in maintaining tissue integrity.

\end{abstract}

\maketitle

\section{Introduction}

Epithelial tissues serve as protective barriers for many organisms, covering organs and forming the outer layer of the skin. To maintain their integrity, these tissues must withstand significant mechanical stresses \cite{Ladoux2017}, otherwise they fracture \cite{Casares2015}, forming holes, which can lead to associated diseases \cite{Vlahakis2003}. A key mechanism by which epithelial tissues preserve their integrity is through coordinated cell motion \cite{Boocock2021, Friedl2009, Rorth2009, Sarate2024, Kreeger2007, DiMeglio2022}, mediated by intercellular contacts. These contacts form networks that transmit local mechanical forces across the tissue \cite{Saraswathibhatla2021}, generating stress patterns that regulate processes such as cell division \cite{Campinho2013}, extrusion \cite{Eisenhoffer2012}, and migration \cite{Collinet2021}.

In some cases, the formation of holes in epithelia can be a crucial step in development. For instance, in multicellular organisms like \textit{Trichoplax adhaerens}, collective movements of the epithelial sheet can drive spontaneous ruptures, serving as a unique mechanism of asexual reproduction \cite{Prakash2021}. However, understanding such phenomena still poses significant challenges, particularly in \textit{in vivo} settings where imaging and mechanical characterization are limited \cite{Roca-Cusachs2017}. As a result, \textit{in vitro} systems have become essential tools for investigating collective cell behaviour under controlled mechanical environments \cite{GomezGonzalez2020, Angelini2010, Murrell2011, Vazquez2022}. 

Recent work on MDCK cell monolayers cultured on polyacrylamide gel substrates has revealed that substrate stiffness strongly influences tissue integrity \cite{Sonam2023}. In particular, tissues cultured on soft substrates can spontaneously form holes, whereas stiffer substrates preserve tissue confluency. The authors of that work highlight how active nematic theories \cite{Doostmohammadi2018, Simha2002} can describe the collective motion of cells and the formation of topological defects (in the cell shape) before hole formation. These defects give rise to local gradients in stresses that can lead to rupture. Other work has similarly highlighted how the rheological properties of the substrate can influence cell-substrate adhesions which, in turn, affect the collective behaviour of cells \cite{Zheng2017}. These observations provide compelling evidence for the role of the supporting structure in mediating the collective behaviour in cell monolayers, and in driving spontaneous hole formation.

Here we use an active nematic multi-phase field model that is able to capture the collective motion observed in cell monolayers to investigate spontaneous hole formation. Building on prior work \cite{Zhang2023}, we represent each cell as a deformable phase field, which evolves under internal passive forces, mechanical interactions with neighbouring cells, and internally generated active stresses arising from actomyosin contractility \cite{Prost2015}. Inspired by recent advances \cite{Fu2024, Rozman2025, Chiang2024a, Chiang2024b, Barret2025, Chaoyu2024}, we introduce internal dissipation, via cell-cell friction, leading to long-range correlations and collective motion, commonly seen in cell monolayers \cite{Andersen2025, Saw2017, Nejad2024}. We demonstrate that, in this regime, spiral patterns in the velocity field emerge as a consequence of antiparallel alignment and motion of two $+1/2$ topological defects in the nematic director field. As the defects move apart, they generate large local strains that can initiate spontaneous hole formation. Our results show in particular that substrate friction modulates this behaviour: high friction suppresses long-range correlations, similar to continuum theories \cite{Krajnc2021, Thijssen2020, Ardaseva2022}, inhibiting hole formation, while reduced friction increases the likelihood of rupture. We also observe, in agreement with experiments, that, as the hole nucleates, local flows align cells along its border, thereby promoting further growth.

We highlight how hole formation, driven by large strain localization near topological defects, follows similar statistics as observed in other disordered systems \cite{Morankar2022, Cipelletti2020}. Due to activity, holes are highly dynamical and can heal. We identify a transition between regimes dominated by short-lived transient holes and those characterised by long-lived persistent rupture, which is governed by the interplay between activity and substrate friction. Our findings offer new insights into how the mechanical support, via a substrate, regulates collective cell motion and mediates cell monolayer integrity.

\section{Model}

We use a 2D multi-phase field model where each cell is described by an independent scalar phase field, $\phi_i(\textbf{x})$, that continuously varies from $0$ to $1$ \cite{Mueller2019, Basan2013}. The motion of each phase field is governed by a local velocity field $\textbf{v}_i(\textbf{x})$ according to the equation of motion

\begin{equation}
\label{eqEOM}
    \partial_{t} \phi_i(\textbf{x}) + \textbf{v}_i(\textbf{x}) \cdot \nabla \phi_i(\textbf{x}) = -J_0 \frac{\delta \mathscr{F}}{\delta \phi_i(\textbf{x})} \\ ,
\end{equation}
 
\noindent where $\mathscr{F}$ is a free energy. The right-hand side of Eq.~\eqref{eqEOM} corresponds to the relaxation dynamics of the cells to a free-energy minimum at a rate $J_0$.

We recover the local velocity field by solving the force balance equation:

\begin{equation}
\label{eqforce}
    \xi_{s} \textbf{v}_i(\textbf{x}) + \xi_{cell} \sum_{\textbf{x}^\prime\in \mathcal{N}(\textbf{x})} \sum_{j\in \textbf{x}^\prime} [\textbf{v}_i(\textbf{x}) - \textbf{v}_j(\textbf{x}^{\prime})] = \textbf{f}_i^{p}(\textbf{x}) + \textbf{f}_i^{a}(\textbf{x}) \\ .
\end{equation}
The first term on the left-hand side of Eq.~\eqref{eqforce} captures the dissipation due to friction from the relative motion of the cell on a substrate with a friction coefficient $\xi_s$. The second term represents the dissipation from the relative motion between neighbouring cells and inside a given cell, with a corresponding friction coefficient $\xi_{cell}$. $\mathcal{N}(\textbf{x})$ are the eight lattice sites surrounding  \textbf{x}, and $j$ corresponds to all phase fields at the lattice site \textbf{x}$^\prime$. Given that the lattice size is normalised to one, this term resembles a Laplacian of the velocity field at the lattice site \textbf{x}, resulting in a viscosity-like term similar to continuum models \cite{Doostmohammadi2018}. Recent work introduced a similar term to a multi-phase field model but with velocities at the level of the individual cell \cite{Chiang2024a, Chiang2024b}. Our implementation allows for spatial variations in velocity within individual cells, thereby enabling the description of nematic activity. In the following, we fix $\xi_{cell} = 1$ and vary $\xi_s$.

The right-hand side of Eq.~\eqref{eqforce} includes passive and active force densities. The passive force density, 

\begin{equation}
\label{pforce}
    \textbf{f}_i^{p}(\textbf{x}) = -\nabla \bigg(\frac{\delta \mathscr{F}}{\delta \phi_i(\textbf{x})} \bigg) \\,
\end{equation}

\noindent represents the passive drive to the minimum of the free energy $\mathscr{F}$, through changes in the cell shape. $\mathscr{F}$ includes a Cahn-Hilliard term that encourages $\phi_i$ to be $1$ inside and $0$ outside the cell $i$, a soft constraint on the area of each cell, and a repulsion energy that penalizes overlaps. Explicit expressions for these terms are given in Sec.~A of the Supplementary Information.

The active force density,
\begin{equation}
\label{activeforce}
    \textbf{f}_i^{a}(\textbf{x}) = -\zeta \textbf{Q}_i \cdot \nabla \phi_i(\textbf{x}) \\,
\end{equation}

\noindent represents the active driving of the cell due to actomyosin activity. The strength of the activity is given by $\zeta$, and $\textbf{Q}_i$ is the nematic tensor that controls the direction of the active forces for each cell $i$. For simplicity, we assume that the nematic tensor follows the deformation tensor of the cell,

\begin{equation}
\label{Qtensor}
    \textbf{Q}_i = - \int dx \bigg[ \nabla \phi_i \nabla \phi_i^T - \frac{1}{2} \text{Tr}(\nabla \phi_i \nabla \phi_i^T) \bigg] \\,
\end{equation}

\noindent so that the eigenvectors of this tensor, $\textbf{d}^{\parallel}$ and $\textbf{d}^{\bot}$, lie along and perpendicular to the elongation of the cell, respectively, and we normalize $\textbf{Q}_i$ by $0.5\sqrt{Q_{xx}^2+Q_{xy}^2}$. Previous works have shown that the collective motion of cell monolayers resembles that of an extensile active nematic system \cite{Balasubramaniam2021, Saw2017}. Thus, we focus on the extensile regime, $\zeta>0$. See Sec.~A of the Supplementary Information and Refs.~\cite{Zhang2020, Zhang2023} for more details of the model.

In the following, we consider a system of $N=100$ deformable cells, with a radius of $R=8$ in lattice units, which move on an underlying periodic square lattice of side $L=140$. These choices lead to packing fractions above $95\%$ meaning that the cells are confluent, tightly packed with no gaps between them. We choose parameters, which are detailed in Sec.~A of the Supplementary Material, to target the low hole density regime, where the typical distance between holes in the monolayer is much greater than the holes' characteristic size, so that it is reasonable to discount hole-hole interactions. The equations of motion~\eqref{eqEOM}--\eqref{Qtensor} are integrated using a finite difference method in space and a predictor-corrector method in time \cite{Mueller2019}. We initialize the system by positioning the cells randomly in the simulation box with a starting radius of $R/2$. We then let the system relax without activity and cell-cell friction for $10000$ time-steps, after which we simulate with activity and cell-cell friction for another $100000$ time-steps. 

Based on typical values for MDCK cells, an average radius of approximately \( 10\,\mu\text{m} \), velocity around \( 20\,\mu\text{m}/\text{h} \), and pressure on the order of \( 100\,\text{Pa} \), as measured via Particle Image Velocimetry and Traction Force Microscopy~\cite{Saw2017}, we estimate that the simulation units correspond to physical units of \( \Delta x \sim 1\,\mu\text{m} \), \( \Delta t \sim 5\,\text{s} \), and \( \Delta F \sim 1.5\,\text{nN} \) for length, time, and force, respectively.

  \begin{figure*}[t]
	\includegraphics{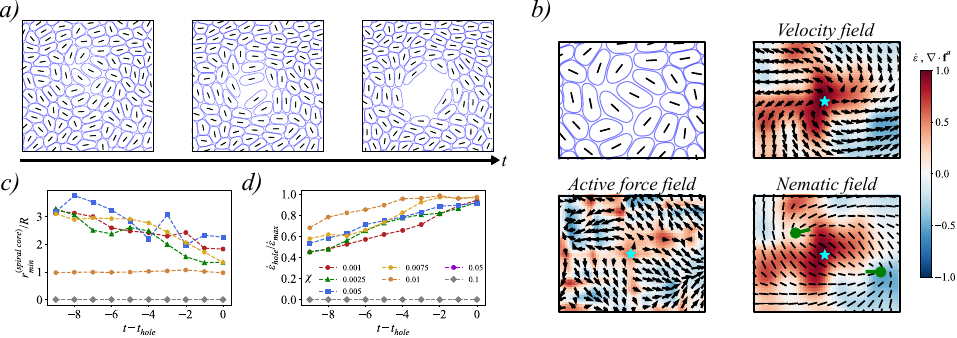}
	\caption{\label{fig1} {\bf  A model cell monolayer during hole formation.} a) Snapshots showing the evolution of the system as a hole nucleates in the monolayer. The boundary of the cells are drawn at $\phi_i(\textbf{x})=0.5$ and the black line inside each cell corresponds to the elongation axis. In b) we zoom in on the area where hole nucleation occurs and image various properties of the monolayer: velocity, active force, and nematic fields. In the velocity field, the cyan star corresponds to the location of a spiral core (also represented in the other fields for reference). In the nematic field, the green dots represent the location of $+1/2$ topological defects with a line pointing towards the tail of the defect. The colours in the velocity and nematic fields represent the local normalised isotropic strain rate calculated from the velocity field. In the active force field, the colours represent the normalised magnitude of the divergence of the active force. The colour bar highlights these normalised quantities for the respective figures. All fields are smoothed using a sliding window of size $2R \times 2R$. c) Distance from the hole's centroid to the closest spiral core, normalised by the cell radius $R$, as a function of the number of time-frames before hole nucleation, where $0$ corresponds to a hole nucleation event. d) Normalised maximum strain rate in an area of radius $R/2$ around the location of hole nucleation as a function of the number of time-frames before nucleation. The normalization is per time-frame, where each time-frame spans $1000$ time-steps. These results were averaged over $10$ simulations with $100$ time-frames. In c) and d), the points at zero distance and strain rate, respectively, represent simulations which do not nucleate holes ($\chi=[0.1, 0.05]$). For $\chi=0.01$ the number of holes formed is small, resulting in poor statistics that artificially flatten the curve’s behaviour.
	}
\end{figure*}

\section{Results}

To focus on the effects of substrate friction and activity on the collective behaviour of cell monolayers it is useful to define $\chi = \xi_s / \xi_{cell}$ as a measure of the hydrodynamic screening length \cite{Krajnc2021, Thijssen2020, Ardaseva2022}. When $\chi \sim 1$, the dissipation is dominated by substrate friction, and the characteristic length scale of velocity and nematic correlations is smaller than a typical cell size. In this regime, the motion of the cells is random, and no holes form in the monolayer. When the substrate friction is sufficiently low that the characteristic length scale of correlations is larger than the cell size (see Supplementary Material Sec.~B), and the activity is large enough to compete with the elastic response of the monolayer, the system enters a regime similar to active turbulence \cite{Doostmohammadi2018}, and holes spontaneously form. This observation underscores how cell-cell friction can be a key physical ingredient to accurately realize monolayer collective behaviour.

Figure \ref{fig1}a) shows a schematic with sequential snapshots of the model cell monolayer during a typical hole nucleation event. Figure \ref{fig1}b) represents the velocity, force, and nematic fields just before hole nucleation. We observe that close to a nucleation event, the cells are arranged in a specific pattern, which is characterised by two $+1/2$ topological defects in the nematic field moving in the vicinity of each other in an antiparallel relative orientation. This correlated motion leads to a net active force pointing outwards from the axis of the defect rotation and the formation of a spiral in the velocity field (point where the divergence of the velocity field is positive). Since the monolayer is compressible, if the active forcing and flows are large enough to overcome the elastic (passive) response that keeps the monolayer confluent, a hole can spontaneously form. We confirm the role of pairs of $+1/2$ topological defects in Sec.~C of the Supplementary Information, by looking at the distance from the hole to the closest defects.

We show that this spatial arrangement is crucial by plotting, in Fig.~\ref{fig1}c) and d), the distance from the hole's centroid to the closest spiral core in the velocity field and the maximum isotropic strain rate, respectively, at the moment of hole formation, $t - t_{hole}= 0$. The \textit{x}-axis of these plots then corresponds to the number of time-frames before the nucleation event (each time-frame corresponds to $1000$ time-steps). The results suggest that, just before the hole forms, both the peak strain in the monolayer and the nearest spiral core localise near the nucleation point (within approximately one cell diameter). These observations highlight how collective motion leads to the localization of strains, driven by active forces, culminating in rupture.

If the activity is too low, the hole is not able to form. This is because the active forces are not large enough to overcome the energy barrier that keeps the monolayer confluent. To confirm this we checked that as the surface tension of the cells increases, thus making cells less deformable and increasing the effective elastic response of the monolayer, the probability of forming a hole decreases (see Supplementary Information Sec.~D).

\begin{figure}[t]
	\includegraphics{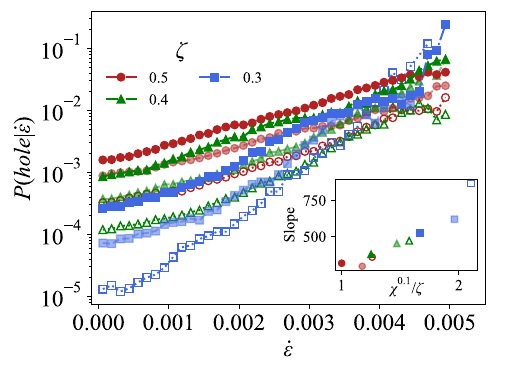}
	\caption{\label{fig2} {\bf Conditional probability of forming a hole for a given strain rate} as a function of local isotropic strain rate and activity (empty symbols correspond to $\chi=0.01$, partially transparent symbols to $\chi=0.005$, and full symbols to $\chi=0.001$). The main plot shows a log-linear scale to highlight the exponential form. The strain rate field is calculated from the divergence of the velocity field. Each strain rate point corresponds to $\dot{\varepsilon}+\Delta\dot{\varepsilon}$, where $\Delta\dot{\varepsilon}=0.000125$. The inset shows the coefficients of the exponential fits as a function of $\chi^{0.1}/\zeta$. Each distributions is averaged over $100$ different simulations with $100$ different time-frames.
    }
\end{figure}

Given that the passive response keeps the monolayer confluent, the strain generated by the active forces needs to be large enough to overcome this energy barrier. Therefore, not all spiral patterns lead to holes, only those that localise enough strain. Figure~\ref{fig2} shows the conditional probability of forming a hole given a local isotropic strain rate, $P(hole|\dot{\varepsilon})$. We measure this by first calculating the strain rate field across the monolayer, for each time-frame. We then count $N_{\dot{\varepsilon}}$, the total number of lattice sites across all time frames and all simulations for which the corresponding strain rate is within a given interval $\dot{\varepsilon} + \Delta \dot{\varepsilon}$. For each strain rate interval, we also check how many of the counted lattice sites open into a hole in the subsequent time-frame, denoted as $ N_{hole, \dot{\varepsilon}}$. Thus, $P(hole|\dot{\varepsilon}) = N_{hole, \dot{\varepsilon}}/N_{\dot{\varepsilon}}$. Note that if a hole grows above a threshold size, corresponding to half of a single cell area, lattice sites are not counted towards either $N_{hole, \dot{\varepsilon}}$ or $N_{\dot{\varepsilon}}$ until the hole closes, as the system size cannot accommodate more than one hole at the same time.

The main plot in Fig.~\ref{fig2} shows that $P(hole|\dot{\varepsilon})$, for different activities and substrate frictions, is well fitted by an exponential. This confirms that high strain rates are exponentially more likely to form a hole, as cells need to generate enough force to overcome the passive response. We remark that the coefficient of the exponential decreases with activity and increases with substrate friction (see inset of Fig.~\ref{fig2}). Higher activity or lower substrate friction generate larger strain rates, but the fraction of the large strain rates that nucleate holes decreases. We hypothesise that this is because there are competing regions of high strain rate for higher activities and lower substrate frictions, so multiple regions of the monolayer may experience large strain rates. However, since steric constraints mean that only one hole can form at a time, only one of the high strain rate regions gives rise to a hole. The inset in Fig.~\ref{fig2} plots the coefficient of the different exponentials against the ratio $\chi^{0.1}/\zeta$, collapsing the points onto the same curve.

Defect-driven fracture is reminiscent of other soft disordered materials like glasses and polymer networks \cite{Morankar2022, Cipelletti2020}, which are highly heterogeneous and whose fracture, when under externally applied stress, is initiated by the accumulation of stresses near defects in the material. However, unlike passive systems, cells are highly motile and active. This means that the monolayer can rupture without any externally applied stresses \cite{Ray2018, HerrmannRoux1990}; the collective motion of the cells can spontaneously generate local stresses (and strains) that lead to holes. Moreover, rupture is a more dynamical process, where holes can grow, shrink, annihilate, and nucleate again at a later time.

\begin{figure*}[t]
	\includegraphics{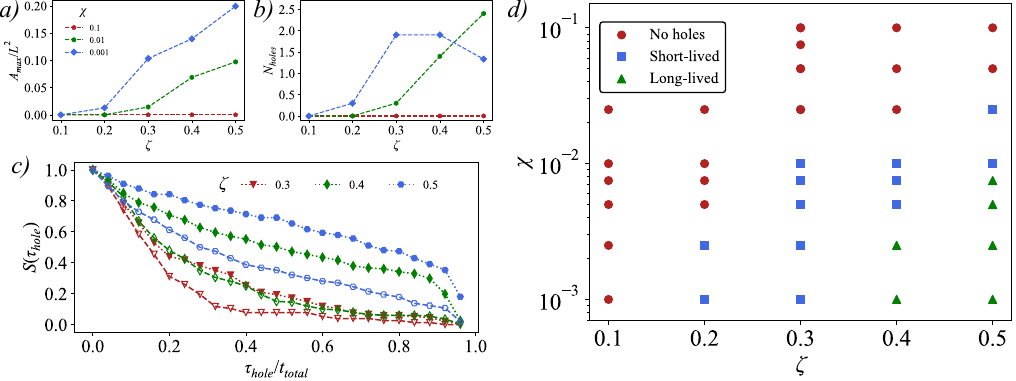}
	\caption{\label{fig3} {\bf Hole lifetime diagram.} a) Normalised maximum hole area and b) total number of holes integrated over the whole simulation as a function of activity for different substrate frictions ($\chi = [0.1, 0.01, 0.001]$). c) Survival probability of holes, i.e, the probability that a hole survives a fraction, $\tau_{hole}/t_{total}$, of the full simulation time, $t_{total}$, for different activities ($\zeta = [0.3, 0.4, 0.5]$) and substrate frictions (empty symbols correspond to $\chi=0.005$ and full symbols to $\chi=0.001$). We fit an exponential to each curve to calculate the characteristic hole lifetime. d) Diagram classifying parameters corresponding to no holes (red circles), only short-lived holes (blue squares), and both short and long lived holes (green triangles). The results in a), b), and d) were each averaged over $10$ simulations with $100$ time-frames. The results in c) were each averaged over $100$ simulations with $100$ time-frames.
	}
\end{figure*}

Higher activity helps generate the forces required to overcome the monolayers' elastic response. Notably, lower substrate friction enhances correlations, enabling velocity field patterns that nucleate holes and stabilize them by aligning cells tangentially along the tissue–hole interface (see Fig.~\ref{fig1}a). This alignment is important since it will make active forces generated by the cells at the interface act along the radial direction of the hole and counteract the opposing passive forces. For the hole to shrink, this alignment needs to be broken by neighbouring cells. Increasing activity or decreasing substrate friction promotes cell alignment, thereby improving hole stability. This behaviour resembles active anchoring \cite{Blow2014}, acting as a mechanism that facilitates both the stabilization and growth of holes.

Figures~\ref{fig3}a) and b) characterise the maximum area and total number of holes as a function of activity and substrate friction, respectively. As either the activity increases or substrate friction decreases, the maximum area of holes increases as the active forces become more able to counteract the passive forces in the monolayer that promote confluency.

Conversely, the total number of holes integrated over the whole simulation varies non-monotonically with activity and substrate friction. This occurs because the typical lifetime of holes increases with increasing activity and decreases with increasing substrate friction. For low activities (or high substrate friction) no holes are able to form. For higher values of activity, holes can nucleate and their lifetime increases as activity increases (or substrate friction decreases), eventually reaching a lifetime comparable to the total simulation time. Since only one hole can form at a time, long lifetimes prevent new ones from forming. This is quantified in Fig.~\ref{fig3}c) where we plot the survival probability of the holes, i.e., the probability that a hole survives $\tau_{hole}$ units of time after nucleation. Note that, above a given value of activity, there are holes that are able to last the entire duration of the simulation. 

The survival probabilities in Fig.~\ref{fig3}c) are well fitted by an exponential distribution and so we can extract the typical time-scale of hole lifetime. If this time-scale is larger than a given threshold, $\tau_t>0.5\ t_{total}$, where $t_{total}$ is the full simulation time, then we consider that the probability of forming persistent holes is non-negligible. Figure \ref{fig3}d) shows a diagram which further illustrates how activity and substrate friction control the formation and persistence of holes. In physical units, \( \tau_t \sim 5\,\text{h} \), which is consistent with the onset of long-lived holes measured in experiments \cite{Sonam2023}.

\section{Discussion}

We develop an active nematic multi-phase field model, with both substrate and internal dissipation, to study hole formation in cell monolayers. We show that holes only form spontaneously when velocity correlations reach beyond the typical size of a single cell. The hole formation is driven by the relative motion of pairs of $+1/2$ topological defects in the nematic field generating spiral velocity patterns, which serve as nucleation sites for the holes. The collective motion leads to local stretching of the monolayer, triggering rupture. Once formed, the resulting flows tend to align cells tangentially along a hole’s edge, promoting radial expansion. However, particularly for lower activities, the alignment is often disrupted by competing cell flows so that holes have a finite lifetime.

We quantify the rupture statistics by measuring the probability of forming holes for a given local strain rate, which follows an exponential form. This defect-driven fracture process is similar to other soft disordered systems \cite{Morankar2022, Cipelletti2020} which are characterised by their high heterogeneity and stress localisation near material defects. However, here, because a cell monolayer is active, holes can be highly dynamic with a lifetime that can be increased by reducing substrate friction or increasing activity.

In the friction dominated regime ($\chi \sim 1$), no holes form spontaneously. Previous work which considered this regime found small gaps in simulations of a multi-phase field model. However, these are typically smaller than the cell size and are driven by a lower effective packing fraction because activity drives elongation of the cells which reduces the effective occupied area \cite{Ardaseva2022}. In our simulations this is not the case; when no hole is present the packing fraction is always above $95\%$ and, when a hole is formed, no other gaps appear in the monolayer. This is strong evidence that the hole nucleation process in our model is indeed due to collective motion and localization of stresses around the nucleation area, and not due to the lowering of the effective packing fraction or de-wetting of the monolayer (as seen in epithiloid tissues \cite{Lv2024}). In Sec.~D of the Supplementary Information we corroborate this by showing that when we increase the elasticity of the monolayer (through the surface tension of the cells) the probability of nucleating holes decreases. 

Long-range flows underlie multiple processes in biology, from the formation of the primitive streak in the chicken embryo \cite{SerranoNajera2020} to glioblastoma tumour progression \cite{Comba2022}. Incorporating internal dissipation into the multi-phase field approach used here has proven essential to generate the flow correlations \cite{Fu2024, Rozman2025} needed to create holes. Moreover, for some simulations, a hole can grow up to $20\%$ of the monolayers' size which suggests that compressibility effects, which are typically not included in continuum theories \cite{Thampi2016}, are necessary to reach a complete description of biological systems.

Recent work on MDCK cells has shown a similar hole nucleation process mediated by the substrate, in which holes in the monolayer nucleate on soft substrates but not on stiffer ones \cite{Sonam2023}. 
Although Ref.~\cite{Sonam2023} does not explicitly address substrate friction, since cell-substrate adhesions are not measured, the authors report higher traction forces on stiffer substrates, consistent with the higher friction regime in our model. 
They show that the cells move collectively, generating flows and forming $+1/2$ and $-1/2$ topological defects, and that hole formation correlates with the location of the defects, where stress gradients are highest. Moreover, experiments also show alignment of the cells tangential to the holes' interface.

However, there are differences between our model and these experiments. We find that, to facilitate local stretching of the monolayer, two $+1/2$ defects oriented antiparallel to each other are needed to create the spiral flows that drive hole nucleation, and it would be interesting to investigate defect-defect correlations in the experiments. Defects leading to spiral patterns in the cell velocity field have also been shown to modulate collective flows in bacterial colonies \cite{Meacock2021} and influence developmental and regenerative processes in Hydra \cite{Maroudas‑Sacks2025}.

Also, we do not see hole formation at $-1/2$ defects in the simulations. We hypothesize that this is due to not taking anisotropic effects into account. The cell monolayer experiments suggest that cell divisions also play a relevant role in hole formation \cite{Sonam2023}, and since this process is anisotropic it could influence the flows around $-1/2$ defects. Hole formation at $-1/2$ defects in bacterial colonies \cite{Copenhagen2021} has also been connected to anisotropic friction effects. Introducing anisotropic friction will be an interesting topic for future research. Another outstanding question is whether modelling the activity using random polar \cite{Vafa2021, Killeen2022}, rather than nematic, forces leads to any changes in behaviour.

We further comment that cell monolayers can serve as a useful biological model for applying ideas from the statistical physics of fracture to biological systems. Previous studies have examined how these systems respond to external stresses and how their elasticity changes with the rate of deformation \cite{Chen2022, Bonfanti2022, Bidhendi2023}, sometimes leading to effects like strain stiffening \cite{Duque2024}. Our results indicate that active systems can spontaneously form defects and generate stresses that affect their own integrity. This dynamic behaviour suggests the need to extend the definition of failure in such materials and highlights the importance of healing \cite{Papafilippou2025, Kreeger2007, Martin2009, Sarate2024} as part of their mechanical response.

\begin{acknowledgments}
DEPP acknowledges support from the UKRI Horizon Europe Guarantee MSCA Postdoctoral Fellowship No. EP/Z002761/1. JR and JMY acknowledge support from the UK Engineering and Physical Sciences Research Council (Award EP/W023849/1) and the ERC Advanced Grant ActBio (funded as UKRI Frontier Research Grant EP/Y033981/1).
\end{acknowledgments}

\renewcommand\thefigure{S\arabic{figure}}
\setcounter{figure}{0}  

\numberwithin{equation}{subsection}
\renewcommand\theequation{S\arabic{equation}}

\section{Supplementary Information}

\subsection{Simulation details}
\label{SMMethods}

We simulate Eqs.~(1)-(2) in the main text using a finite-difference scheme on a square lattice with a predictor-corrector step. We also use a space decomposition method, as in \cite{Mueller2019}, where each phase field is only solved inside a subdomain that covers an area of $30\times 30$ lattice sites, centred on the centre of mass of the phase field. This subdomain is variable after the equilibration protocol. We set a threshold value of $\phi_i(\textbf{x})=0.01$, and if the distance from \textbf{x} to the sides of the subdomain is smaller than $4$ lattice sites, then we increase the subdomain size. On the other hand, if it is larger, we decrease it. We also do not enforce a square subdomain. This optimization helps deal with highly elongated cells without fixing a large subdomain from the start.

The free energy term in Eq.~(1) of the main text defines the dynamics of the individual interfaces and is written as $\mathscr{F} = \mathscr{F}_{CH} + \mathscr{F}_{area} + \mathscr{F}_{rep}$, where

\begin{equation}
\label{CH}
    \mathscr{F}_{CH} = \sum_i \frac{\gamma}{\lambda} \int d\textbf{x} \big[ 4\phi_i^2(\textbf{x}) (1-\phi_i(\textbf{x}))^2 + \lambda^2 (\nabla \phi_i(\textbf{x}))^2  \big] \\,
 \end{equation}

\begin{equation}
\label{area}
    \mathscr{F}_{area} = \sum_i \mu \bigg[ 1 - \frac{1}{\pi R^2} \int d\textbf{x} \phi_i^2(\textbf{x})   \bigg] \\,
 \end{equation} 

\begin{equation}
\label{rep}
    \mathscr{F}_{rep} = \sum_i \sum_{j\neq i} \frac{\kappa}{\lambda} \int d\textbf{x} \phi_i^2(\textbf{x}) \phi_j^2(\textbf{x}) \\.
\end{equation}

\noindent Equation \eqref{CH} is a Cahn-Hilliard free energy that encourages $\phi_i(\textbf{x})$ to be equal to $1$ or $0$ inside or outside cell $i$, respectively. The cell boundary is located at $\phi_i(\textbf{x}) = 1/2$ and has width $\sim 2\lambda$, set by the gradient term. $\gamma / \lambda$ sets an energy scale. This term controls the deformability of the individual cells. Cell compressibility is described by Eq.~(\ref{area}), which imposes a soft constraint, of strength $\mu$, restricting the area of each cell to $\pi R^2$. Equation~(\ref{rep}) penalizes overlap between cells with an energy scale $\kappa / \lambda$.

We simulate the dynamics of cells of radius $R = 8$ in a periodic domain. Initially, cells with radius $R/2$ are placed randomly. They are then relaxed for $10000$ time-steps under passive dynamics without cell-cell friction to reach confluence. The simulations are then run for $100000$ time-steps and the state of the simulation is printed every $1000$ time-steps. The lattice spacing is set to $\Delta x = 1$ and the interval between time-steps is set to $\Delta t = 0.1$. Parameter values are $\lambda = 2.0$, $\gamma = 0.06$, $\kappa = 0.5$, $\mu = 20$, and $J_0 = 1$.

The monolayers' coarse-grained fields can be calculated at each lattice point using 

\begin{equation}
\label{fields}
    A(\textbf{x}) = \sum_i a_i(\textbf{x}) \phi_i(\textbf{x}) \\,
\end{equation}

\noindent where $a_i(\textbf{x})$ corresponds to a given measure at the lattice site \textbf{x}, such as velocity, active force or nematic field. Since the nematic director is given by the elongation of the cell, we consider that $Q_i(x) \equiv Q_i$. We then smooth each individual field to reduce noise using a sliding window of size $2R\times 2R$. A defect is identified at a point \textbf{x} of a smoothed field by examining the value of the field at the 8 neighbouring points of \textbf{x}.
The orientation of the defects is calculated by the method in Ref.~\cite{Vromans2016}. We calculate the isotropic strain rate using the smoothed velocity field, $\dot\varepsilon(\textbf{x}) = \nabla \cdot \textbf{v}(\textbf{x}) / 2$. We consider a hole has nucleated when the void area is above $\pi R^2/4$, and we set a lattice site to be void when the sum of all phase fields on it is below a threshold, i.e., $\sum_i \phi_i(\textbf{x}) < 0.01$.

\subsection{Velocity and nematic director correlations with internal dissipation}

Previous modelling approaches with the multi-phase field model have mostly focused on using substrate friction as the main dissipation contribution \cite{Zhang2020, Zhang2023}. Recently, cell-cell friction has been introduced into this framework, in the context of polar active cells \cite{Chiang2024a, Chiang2024b}. The main difference to our implementation is that here we calculate the velocity field at every lattice point, instead of only at the centroid of the cell. This makes the description proposed in Eq.~(2) of the main texxt closer to a Laplacian in the local velocity field, similar to continuum theories \cite{Doostmohammadi2018}.

\begin{figure*}[t]
	\includegraphics{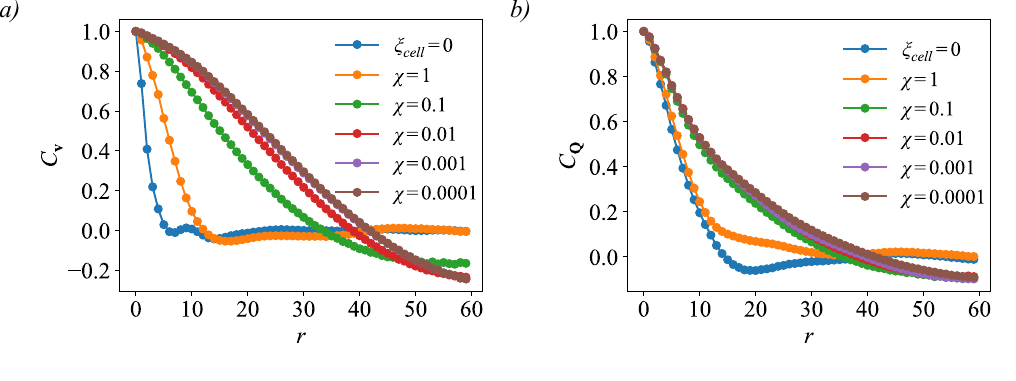}
	\caption{\label{figSM1} Velocity and nematic director correlations as defined by Eqs.~\eqref{Cv} and \eqref{CQ} for different ratios of substrate and cell-cell friction, $\chi = \xi_s / \xi_{cell}$. For the case $\xi_{cell} = 0$, we use $\xi_s = 1$. These results were averaged over $3$ samples with a simulation box of size $120\times 120$.}
\end{figure*}

Here, we highlight the main differences between a \textit{dry} modelling approach ($\xi_{cell} = 0$) and a \textit{wet} modelling approach ($\xi_{cell} > 0$ and $\xi_s\rightarrow0$), as commonly referred in the literature (see Ref.~\cite{Marchetti2013}). Figure \ref{figSM1} shows that in the \textit{dry} model correlations in the velocity and nematic director decay quickly, with a correlation length of the order of the single cell ($r<8$). As we approach the \textit{wet} model regime, these correlations increase as the substrate friction $\xi_s$ is decreased. Similarly to active nematic theories, substrate friction acts as a screening for hydrodynamic effects \cite{Thijssen2020}.

The correlations are calculated through the non-smoothed velocity and nematic fields. We create the respective fields at every lattice point using Eq.~\eqref{fields} and then calculate the correlations for a given snapshot after $100000$ time-steps,

\begin{equation}
\label{Cv}
    C_{\textbf{v}}(r) = \frac{\langle \textbf{v}(\textbf{x}) \cdot \textbf{v}(\textbf{x}^\prime) \rangle}{\langle \textbf{v}(\textbf{x}) \cdot \textbf{v}(\textbf{x}) \rangle} \\,
\end{equation}

\begin{equation}
\label{CQ}
    C_{\textbf{Q}}(r) = \frac{\langle \textbf{Q}(\textbf{x}) : \textbf{Q}(\textbf{x}^{\prime}) \rangle}{\langle \textbf{Q}(\textbf{x}) : \textbf{Q}(\textbf{x}) \rangle} \\,
\end{equation}

\noindent where $r=|\textbf{x}-\textbf{x}^\prime|$, and $\textbf{x}$ and $\textbf{x}^\prime$ represent two lattice sites at distance $r$. To make sure that no holes are formed, we decrease the size of the simulation box to $120\times 120$.

\subsection{Distance from hole nucleation to nematic topological defects}

\begin{figure}[t]
	\includegraphics{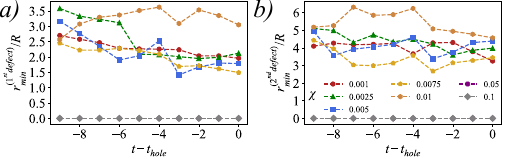}
	\caption{\label{figSM2} a) and b) represent the distance from the hole's centroid to the closest and second closest $+1/2$ topological defects respectively, normalised by the cell radius $R$, as a function of the number of time-frames before hole nucleation, where $0$ corresponds to a hole nucleation event. These results were averaged over $10$ simulations with $100$ time-frames each. The points at zero distance represent simulations which do not nucleate holes ($\chi=[0.1, 0.05]$).}
\end{figure}

As discussed in the main text, hole formation arises from the collective flow of cells near spiral cores. In particular, spiral patterns are formed when two $+1/2$ topological defects in the nematic field move in the vicinity of each other. To investigate this, Fig.~\ref{figSM2} shows the minimum distance between the hole nucleation centroid and the two nearest $+1/2$ defects during the time leading up to nucleation. While the hole does not always nucleate precisely between the defects, it remains consistently closer to two defects, with both defects typically located within approximately two cell diameters. These results support the role of $+1/2$ topological defect pairs in driving hole formation.

\subsection{Surface tension decreases hole nucleation probability}

\begin{figure}[t]
	\includegraphics{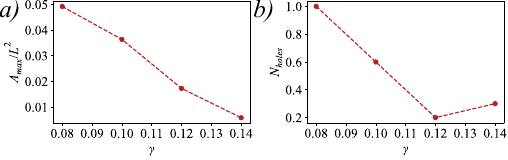}
	\caption{\label{figSM3} a) Normalised maximum hole area and b) total number of holes integrated over the whole simulation as a function of the surface tension of the cells, for $\chi = 0.001$ and $\zeta=0.3$. These results were averaged over $10$ simulations with $100$ time-frames each.}
\end{figure}

In this section, we highlight that hole formation can be viewed as an activity-driven nucleation process, i.e. the active forces of the cells need to overcome the energy barrier imposed by the elasticity of the monolayer. To support this statement, we increase the elasticity of the monolayer through the surface tension of the cells ($\gamma$ in the Cahn-Hilliard free energy term in Eq.~\eqref{CH}). This parameter decreases the deformability of the cells, making the elastic response stronger. Figure \ref{figSM3} shows that as $\gamma$ increases, both the average number and the average maximum area of holes decrease.

\providecommand{\noopsort}[1]{}\providecommand{\singleletter}[1]{#1}%

\end{document}